\documentclass[final,5p,times,twocolumn]{elsarticle}

\usepackage{amssymb}

\usepackage{multirow}
\journal{Nuclear Physics A}
\newcommand{\pt} {\mbox{$p_{\rm t}$}}

\newcommand{\sqrtsNN}{\sqrt{s_{\rm NN}}}
\newcommand {\tev} {\mbox{${\rm TeV}$}}

\begin{document}

\begin{frontmatter}


\title{Upgrade of the ALICE Inner Tracking System}
\author{Stefan Rossegger\corref{cor1}\fnref{label2}}
\ead{stefan.rossegger@cern.ch}
\fntext[label2]{on behalf of the ALICE ITS collaboration}
\address{European Organization for Nuclear Research (CERN), Geneva, Switzerland}




\begin{abstract}
The Inner Tracking System (ITS) is the key ALICE detector for the study of heavy flavour production at LHC. Heavy flavor can be studied via the identification of short-lived hadrons containing heavy quarks which have a mean proper decay length in the order of 100-300~$\mu$m. To accomplish this task, the ITS is composed of six cylindrical layers of silicon detectors (two pixel, two drift and two strip) with a radial coverage from 3.9 to 43~cm and a material budget of 1.1~\% X0 per layer. 

In order to enhance the ALICE physics capabilities, and, in particular, the tracking performance for heavy-flavour detection, the possibility of an ITS upgrade has been studied in great detail. It will make use of the spectacular progress made in the field of imaging sensors over the last ten years as well as the possibility to install a smaller radius beampipe. The upgraded detector will have greatly improved features in terms of: the impact parameter resolution, standalone tracking efficiency at low $p_{t}$, momentum resolution and readout capabilities.

The usage of the most recent monolithic and/or hybrid pixel detector technologies allow the improvement of the detector material budget and the intrinsic spatial resolution both by a factor of three with respect to the present ITS. The installation of a smaller beam-pipe reduces the distance between the first detector layer and the interaction vertex. Under these assumptions, simulations show that an overall improvement of the impact parameter resolution by a factor of three is possible.

The Conceptual Design Report, which covers the design and performance requirements, the upgrade options, as well as the necessary R\&D efforts, was made public in September 2012. An intensive R\&D program has been launched to review the different technological options under consideration. The new detector should be ready to be installed during the long LHC shutdown period scheduled in 2017-2018.
\end{abstract}


\end{frontmatter}



\section{Introduction}

The general upgrade plans for the ALICE experiment, its experimental strategy and the long-term physics 
goals are discussed in \cite{ALICE_LOI}. As described there, future physics topics require a significant
 improvement of the measurements of heavy-flavour hadrons, quarkonia and low-mass dileptons at low transverse
 momenta. Furthermore, the upgraded ALICE detector will be able to make full use of a high-luminosity LHC 
(L = $6 \times 10^{27}$~cm$^{-2}$s$^-1$) for Pb--Pb through the improvement of the readout-rate 
capabilities. The upgraded detector system will allow the inspection of up to 50~kHz Pb--Pb and 200~kHz pp interactions.

The upgrade of the Inner Tracking System (ITS) is an essential ingredient for the improved resolution of the distance of closest approach between a track and the primary vertex (impact parameter resolution), the standalone ITS tracking performance as well as the readout-rate capabilities. The Conceptual Design Report for the Upgrade of the ALICE ITS detector \cite{itsCDR} covers the current status, the ongoing R\&D efforts, the design options, the choice and ongoing tests of the different technological options as well as the expected detector and physics performance.

\section{Current and Upgraded Inner Tracking System}
\label{sec:UpgrScen}

The current Inner Tracking System (ITS) of ALICE is composed of six cylindrical layers of silicon detectors placed around the LHC beam-pipe and covers the central rapidity region ($|\eta|<1$) \cite{ALICEjinst}. Three different technologies are in use: two layers of Silicon Pixel Detector (SPD), two layers of Silicon Drift Detector (SDD) and two layers of double sided Silicon Strip Detector (SSD). The geometrical parameters of the layers (radial position, length along beam axis, spatial resolution) and the material budget are summarized in Tab.~\ref{tab:itsCurrent}. 

The present ITS was designed and optimized mainly to obtain a good track extrapolation to the vertex for tracks reconstructed in the surrounding detector, the Time Projection Chamber (TPC). Therefore, the ITS improves significantly the impact parameter resolution $d_0$ which is a prerequisite for accessing decay channels with a short decay length. Furthermore, the four layers equipped with drift and strip detectors provide a measurement of the specific energy loss ${\rm d} E / {\rm d}x$ and are therefore used for particle identification.

\begin{table}[t]
\caption{Characteristics of the six ITS layers, the beam-pipe and the thermal shields (compare \cite{ALICEjinst}).}
\footnotesize
\begin{tabular}{|p{1.7cm}|c|c|c|c|}
\hline
Layer / Type & $r$ [cm] & $\pm z$ [cm] &
$\begin{array}{c}
{\rm Intrinsic} \\
{\rm resolution \, \left[\mu m \right]} \\
r\phi \quad\quad z 
\end{array}$&
$\begin{array}{c}
{\rm Material} \\
{\rm  budget} \\
 X/X_0 \, \left[\%\right] 
\end{array}$\\
\hline
Beam pipe & 2.94&  -   &   - & 0.22\\
1 / pixel & 3.9 & 14.1 &   $12\quad 100$ & 1.14 \\
2 / pixel & 7.6 & 14.1 &   $12\quad 100$ & 1.14 \\
Th. shield & 11.5 & -   &  - & 0.65\\
3 / drift & 15.0 & 22.2 &  $35\quad25$ & 1.13 \\
4 / drift & 23.9 & 29.7&   $35\quad25$ & 1.26 \\
Th. shield & 31.0 & -   &  - & 0.65\\
5 / strip & 38.0 & 43.1&   $20\quad 830$ & 0.83 \\
6 / strip & 43.0 & 48.9&   $20\quad 830$ & 0.83 \\
\hline
\end{tabular}
\label{tab:itsCurrent}
\end{table}

As described in great detail in \cite[chap.\,3]{itsCDR}, the proposed upgrade of the ITS will make use of the
foreseen reduction of the LHC beam-pipe radius from currently 2.94~cm to 1.98~cm which will allow 
a significant decrease of the distance between the first detector layer and the interaction vertex.
The material budget can be significantly reduced by using ultra-thin hybrid pixel detectors or monolithic 
silicon pixel detectors allowing $X/X_0=0.5$\% or $X/X_0=0.3$\% respectively.
Furthermore, state-of-the-art hybrid pixel detectors would allow smaller pixel sizes of 
e.g. 30$\times$30~$\mu$m$^2$, while monolithic pixel detectors are currently featuring pixel 
sizes of 20$\times$20~$\mu$m$^2$, both significantly smaller than the current size of 50$\times$425~$\mu$m$^2$.

Both semi-analytical and Monte-Carlo methods have been exploited to estimate the transverse momentum (\pt) 
dependence of the impact parameter resolution in azimuth ($r\phi$) and beam ($z$) direction, the \pt~resolution,
 as well as the tracking efficiency by varying the number of layers, their radii, material budget and their 
intrinsic resolutions. 

Different design options are currently being reviewed and compared in terms of space,
 design feasibility, integration, manufacturing costs, as well as standalone tracking efficiency and capability. 
Furthermore, an increase of the read-out rate requires a complete redesign of the current ITS due to the rate
 limitations of the current SDD (read-out time approx. 1~ms). The characteristics of the most promising upgrade 
scenario are given in Tab.~\ref{tab:itsNew}.  

\begin{table}[t!]
\caption{Characteristics of the upgrade scenario. 
The numbers in 
brackets refer to the case of microstrip detectors.}
\centering
\footnotesize
\begin{tabular}{|p{1.7cm}|c|c|c|c|c|c|}
\hline
Layer / Type & $r$ [cm] & $\pm z$ [cm] &
$\begin{array}{c}
{\rm Intrinsic} \\
{\rm resolution \, \left[\mu m \right]} \\
r\phi \quad\quad z 
\end{array}$&
$\begin{array}{c}
{\rm Material} \\
{\rm  budget} \\
 X/X_0 \, \left[\%\right] 
\end{array}$\\
\hline
Beam pipe             & 2.0 &  -   &      -                      &  0.22\\
1 / pixel         & 2.2 & 11.2 & $4\quad\quad 4$  & 0.30  \\
2 / pixel         & 2.8 & 12.1 & $4\quad\quad 4$  & 0.30  \\
3 / pixel         & 3.6 & 13.4 & $4\quad\quad 4$  & 0.30  \\
4 / pixel (strip) & 20.0 & 39.0 & $4\, (20)\quad 4\, (830)$& 0.30 (0.83) \\
5 / pixel (strip) & 22.0 & 41.8 & $4\, (20)\quad 4\, (830)$& 0.30 (0.83) \\
6 / pixel (strip) & 41.0 & 71.2 & $4\, (20)\quad 4\, (830)$& 0.30 (0.83) \\
7 / pixel (strip) & 43.0 & 74.3 & $4\, (20)\quad 4\, (830)$& 0.30 (0.83) \\
\hline
\end{tabular}
\label{tab:itsNew}
\end{table}

\begin{figure}[!t]
\centering
\includegraphics[width=0.8\linewidth]{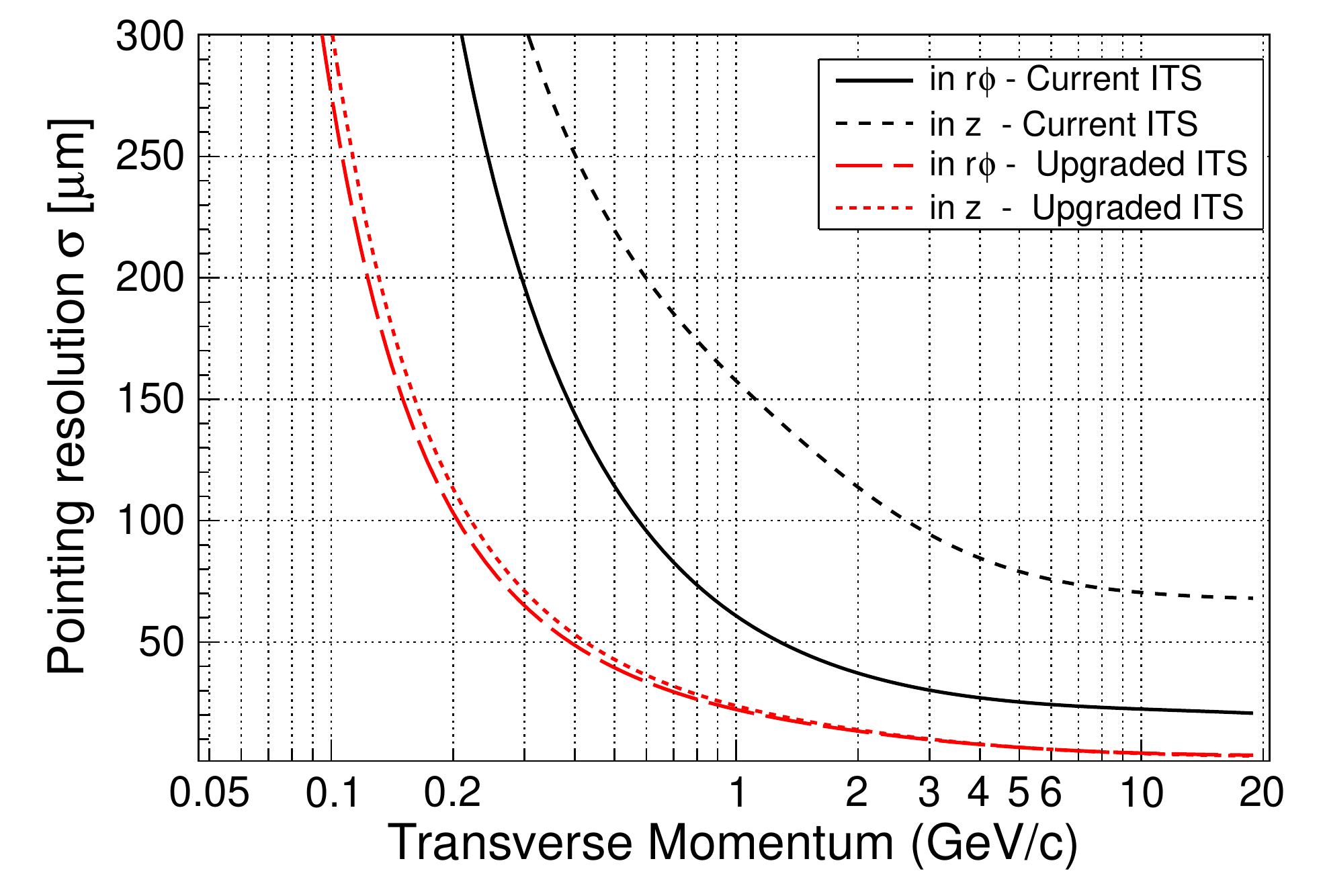}
\caption{Pointing resolution to the vertex of charged pions as a function of the transverse momentum 
for the current ITS and the upgrade scenario.}
\label{fig:resPerformance}
\vspace{6ex}
\centering
\includegraphics[width=0.8\linewidth]{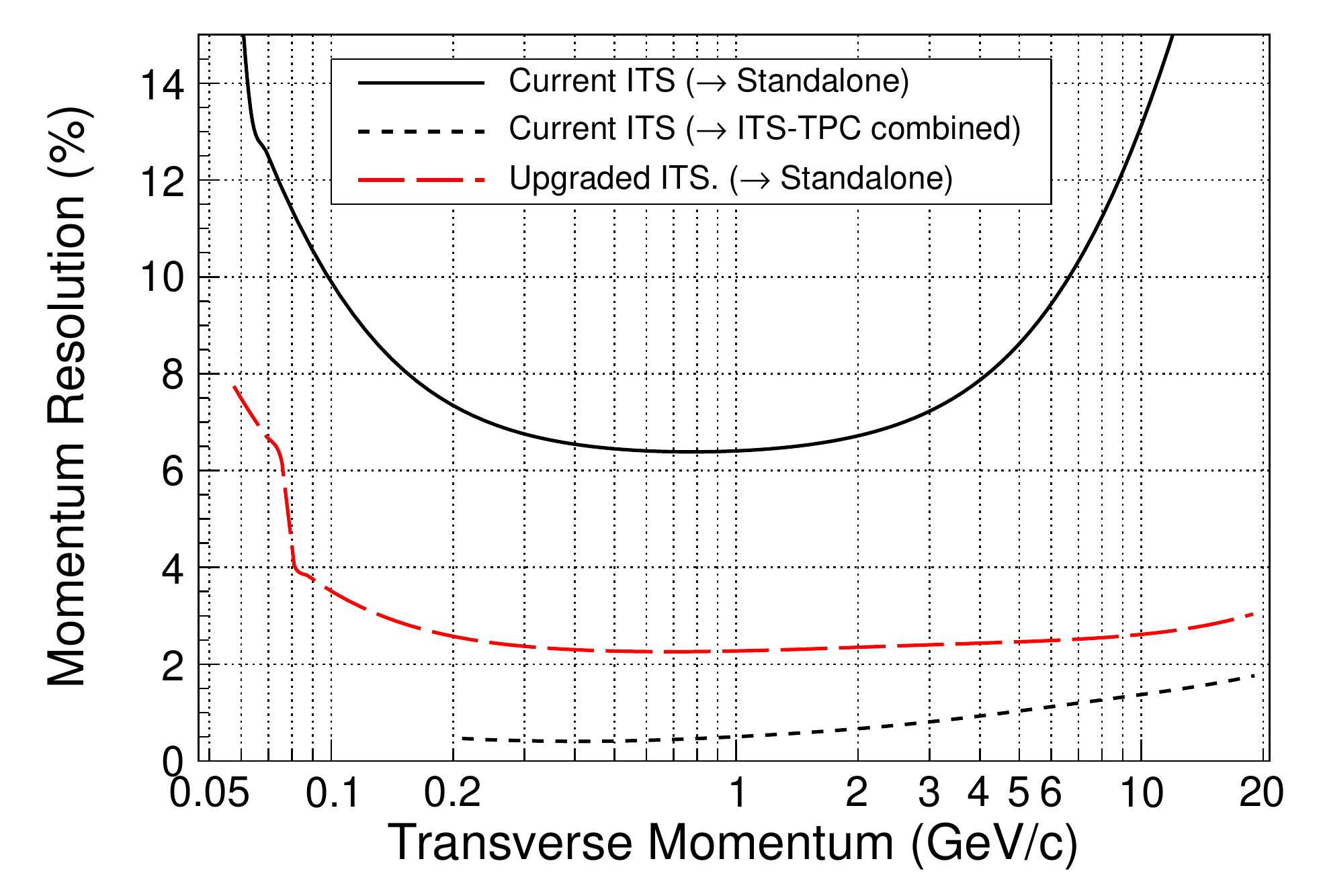}
\caption{Transverse momentum resolution as a function of \pt for charged pions for the current 
ITS and the upgrade scenario.}
\label{fig:ptPerformance}
\vspace{6ex}
\centering
\includegraphics[width=0.8\linewidth]{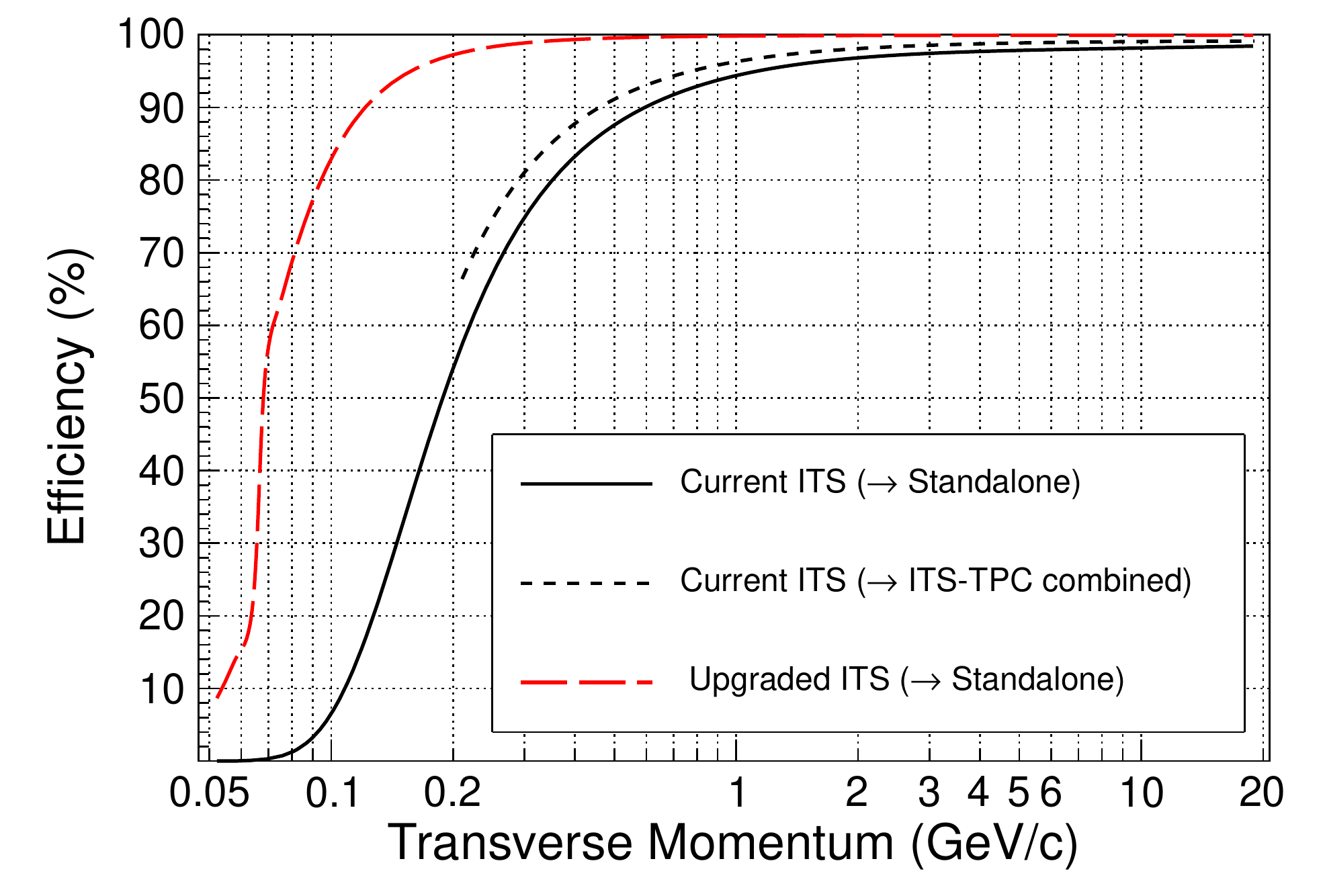}
\caption{Tracking efficiency of charged pions for the current ITS and the upgrade scenario.}
\label{fig:effPerformance}
\end{figure}

The performance of the current ITS is compared against the expected performance of the proposed upgrade 
scenario in the Figs.~\ref{fig:resPerformance},~\ref{fig:ptPerformance} and~\ref{fig:effPerformance} for both, 
the ITS stand-alone and ITS-TPC combined tracking mode. 
Figure~\ref{fig:resPerformance} displays the pointing resolution to the vertex for charged pions. 
Both the $r\phi$ and $z$ components are shown in the plots. The pointing resolution is
 mostly determined by the properties of the first layers.
As an example, at \pt~of about 400~MeV/$c$, an improvement of 3 and 5  is achieved for the $r\phi$ and $z$ 
components respectively. It should be noted that for the present ALICE set-up the ITS-TPC combined 
tracking provides at high \pt~a sizeable improvement with respect to the ITS stand-alone tracking. 
In the case of the upgrade ITS, adding the information of the TPC does not yield any further improvement.

Figure~\ref{fig:ptPerformance} illustrates the improvements in the transverse momentum ($p_{\mathrm{t}}$) 
resolution. For the ITS stand-alone tracking mode, the upgrade scenario yields a dramatic improvement (at least a 
factor 3). The improvement in the ITS-TPC combined tracking mode is due only to the reduction of the 
material budget in the innermost layers. 

Finally, figure~\ref{fig:effPerformance} exhibits a comparison of the tracking efficiency between the current ITS
 and the upgrade scenario in a central Pb--Pb collision at maximal LHC energy 
($\sqrtsNN=5.5~\tev$, ${\rm d}N_{\rm ch} / {\rm d}\eta \simeq  2000$). The stand-alone tracking efficiency
 of the upgrade scenario will surpass the achievable efficiencies of the current ITS design in ITS-TPC combined 
tracking mode. Especially towards lower \pt, a significant enhancement in terms of statistics can therefore be expected.

\section{Technical Implementation}

An R\&D program has been launched to study the various technical options, such as choice of technology
 and architecture. More details can be found in ~\cite[chap.\,4]{itsCDR}. 

As described in the previous section, the new ITS will consist of seven layers of silicon tracking detectors. 
The two options currently under study are:

\begin{itemize}
\item \textbf{Layout 1:} All seven layers are built using silicon pixel detectors.
\item \textbf{Layout 2:} The innermost three layers are equipped with silicon pixel detectors 
             followed by four (outer) layers of silicon strip detectors.
\end{itemize}

Table~\ref{tab:specs_inner} summarizes the technical specifications for the inner and outer layers:

\begin{table}[ht]
\footnotesize
\begin{center}
\caption{Technical specifications for the new ITS layers.}
\label{tab:specs_inner}
\begin{tabular}{|l|c|c|}
\hline
Parameter                 & Inner layers         & Outer layers \\ \hline \hline
Chip Size                 & (15~$\times$~30)~mm & - \\ \hline
Pixel Size (r-$\phi$)     & 20-30~$\mu$m         & $\le$~70~$\mu$m \\  \hline
Pixel Size (z)            & 20-50~$\mu$m          & $\le$~2~cm  \\  \hline
Readout Time              & $\le$~30~$\mu$s       & $\le$~30~$\mu$s \\  \hline
Hit Density               & 150~hits/cm$^{2}$    &  $\approx$1~hit/cm$^{2}$ \\  \hline
Radiation Levels          & 700~krad (TID)       & 10~krad (TID)  \\
       & 1~$\times$~10$^{13}$~n$_{eq}$/cm$^{2}$~(NIEL) &  3~$\times$~10$^{11}$~n$_{eq}$/cm$^{2}$~(NIEL) \\ \hline
\end{tabular}
\end{center}
\end{table}

Two pixel detector technologies are under consideration, namely hybrid and monolithic silicon pixel detectors:

{\itshape Monolithic sensors} embed the sensing diode and front-end electronics on the same silicon chip. They are produced using commercial CMOS
technologies that are optimized for large volumes and are therefore very cost-effective. The sensor can be thinned down to 50~$\mu$m,
offering a significant reduction in material budget. Pixel
sizes of 20~$\mu$m~$\times$~20~$\mu$m or less are standard with these technologies.
The quadruple-well CMOS Imaging Sensor (CIS) process provided
by TowerJazz\footnote{TowerJazz, http://www.jazzsemi.com/}  has been identified as the most promising candidate and has thus
been selected for a dedicated R\&D within ALICE.

One key question to be addressed within such R\&D is the radiation hardness of CMOS sensors in view of the expected radiation levels foreseen in ALICE.
Additional issues are the design of an optimized pixel cell and of a low power readout architecture.

{\itshape Hybrid pixels} are a mature technology presently employed in all four major LHC experiments.
In a hybrid detector, the sensor and the front-end electronics are fabricated on two 
separate chips and then mated with the bump bonding technique. This offers the
advantage of optimizing the sensor and its front-end electronics separately at the
expense of extra thickness and cost. The radiation load foreseen in the ALICE
ITS does not present a concern for this type of sensors.
State-of-the-art hybrid pixels have a cell
size of 50~$\mu$m~$\times$~50~$\mu$m, but the evolution of the interconnection
technology is expected to put cell sizes of 30~$\mu$m~$\times$~30~$\mu$m soon
within reach. In the recent months the focus on the hybrid pixel detectors was on the production 
of assemblies with the final thicknesses and the completion of the production of the first edgeless 
epitaxial silicon sensors. 

Two readout architectures are under study:

The first architecture is based on the {\itshape rolling shutter technique}. 
The time between two consecutive readouts of the same row defines 
the integration time, which is of the order of 100~$\mu$s for state
of the art CMOS sensors. This time must be significantly reduced to 
comply with the 50~kHz interaction rate.
A sensor with an extension of the rolling shutter, based on the low power 
architecture of ULTIMATE~\cite{Valin2012}, is under development for the ITS upgrade. 
This new sensor, called MISTRAL, is intended 
to increase the readout speed by a factor of at least four and 
to improve the radiation tolerance by one order of magnitude.
The estimated power density for these chips is $\lesssim$ 400~mW/cm$^2$.

The second topology employs a {\itshape sparsified readout} of the pixel matrix,
so that only zero-suppressed data are transferred to the periphery. 
For a scenario of 100 tracks per cm$^2$ per event and a cluster multiplicity of eight, 
the readout requires 300~ns. Even considering some inhomogeneity in the hit distribution, 
the transfer of only zero suppressed data from the matrix to the periphery allows a 
full scan in a time on the order of 1~$\mu$s. It has been estimated that the 
power consumption of this priority encoding scheme is less than 30~mW/cm$^2$.

A design and prototyping effort is underway to identify the most suitable readout approach.
The choice between the two schemes will depend on readout speed, 
power consumption, and noise of the prototypes.

\begin{figure}[ht]
\centering
\includegraphics[width=0.98\linewidth]{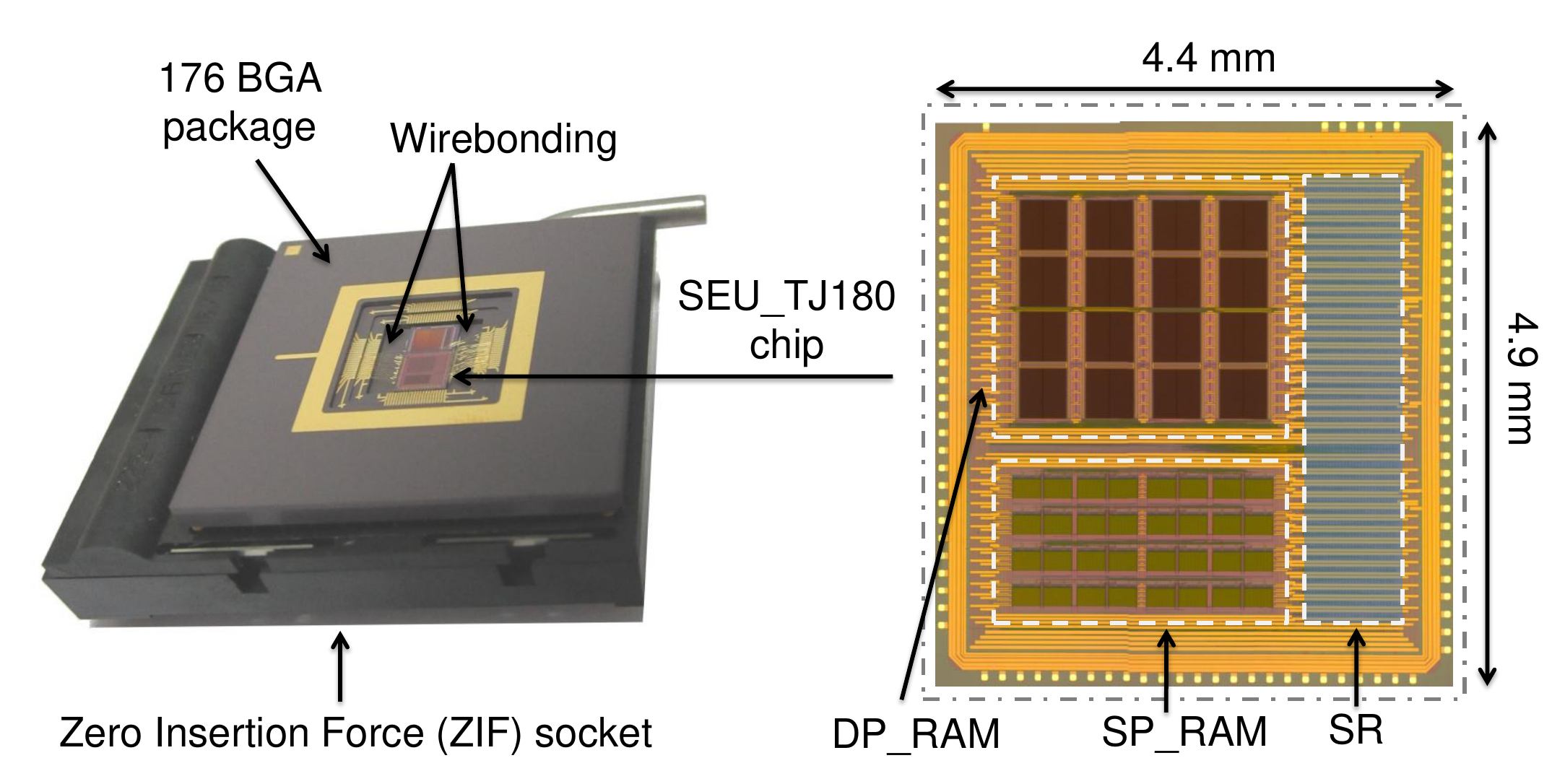}
\caption{Picture of the SEU\_TJ180 structure indicating the dimensions and different blocks (right) and picture of one SEU\_TJ180 structure mounted on the test socket (left).}
\label{fig:SEU_chip}
\end{figure}

To simulate the impact of the expected radiation levels systematic irradiation tests using X-rays, 
protons and neutrons were carried out throughout 2012 and onwards on various sensor, 
analog and digital test structures.
To disentangle the various radiation-induced effects, three types of structures are currently under investigation:

\begin{itemize}
\item Basic structures (diodes and transistors): 2 test structures were designed and implemented in 
TowerJazz 0.18~$\mu$m CMOS technology in order to study basic operational parameters such as 
threshold voltage, transconductance, and dark current as a function of radiation type and dose as 
well as of the layout. Test structures containing various types of NMOS and PMOS transistors and 
capacitors for different epitaxial thicknesses have already been produced in earlier runs 
and are available for radiation tests (RAL structures). Special test structures (TID\_TJ180) 
consisting of various transistor types as well as breakdown diodes were developed and produced 
in the first half of 2012 (ALICE\_ITS\_TJ180). 
\item Digital structures: a test chip for dedicated SEU tests has been designed and implemented in
 TowerJazz 0.18~$\mu$m CMOS technology (SEU\_TJ180) consisting of various RAM structures and shift 
registers to test the stability and reliability of switching states during irradiation with charged 
particles (see Fig.~\ref{fig:SEU_chip}). 
\item Full sensor structures (including analog and digital front end electronics): MIMOSA32 
prototype matrices were produced in TowerJazz 0.18~$\mu$m CMOS technology containing both analog
 and digital elements that allow the characterization of full sensor prototypes under irradiation
 and test beam conditions.
\end{itemize} 


Beam tests were performed in summer 2012 at the CERN-SPS 
with 60-120~GeV particles, which allowed evaluating the 
chip detection performances at two different coolant 
temperatures (T$_c$ =15$^{\circ}$C and 30$^{\circ}$C), 
before and after a combined radiation load of 1~Mrad and 
10$^{13}$~n$_{eq}$/cm$^2$. The study is based on different 
MIMOSA32 dies, tested individually on a beam telescope 
composed of four pairs of microstrip detectors. The detection 
performances (e.g. signal charge collected, pixel noise, 
signal-to-noise ratio (SNR), hit cluster properties, detection 
efficiency) were derived from a total sample of about 50,000 
tracks reconstructed in the beam telescope and traversing 
the pixel array under test.   

\begin{figure}[ht]
\centering
\includegraphics[width=0.8\linewidth]{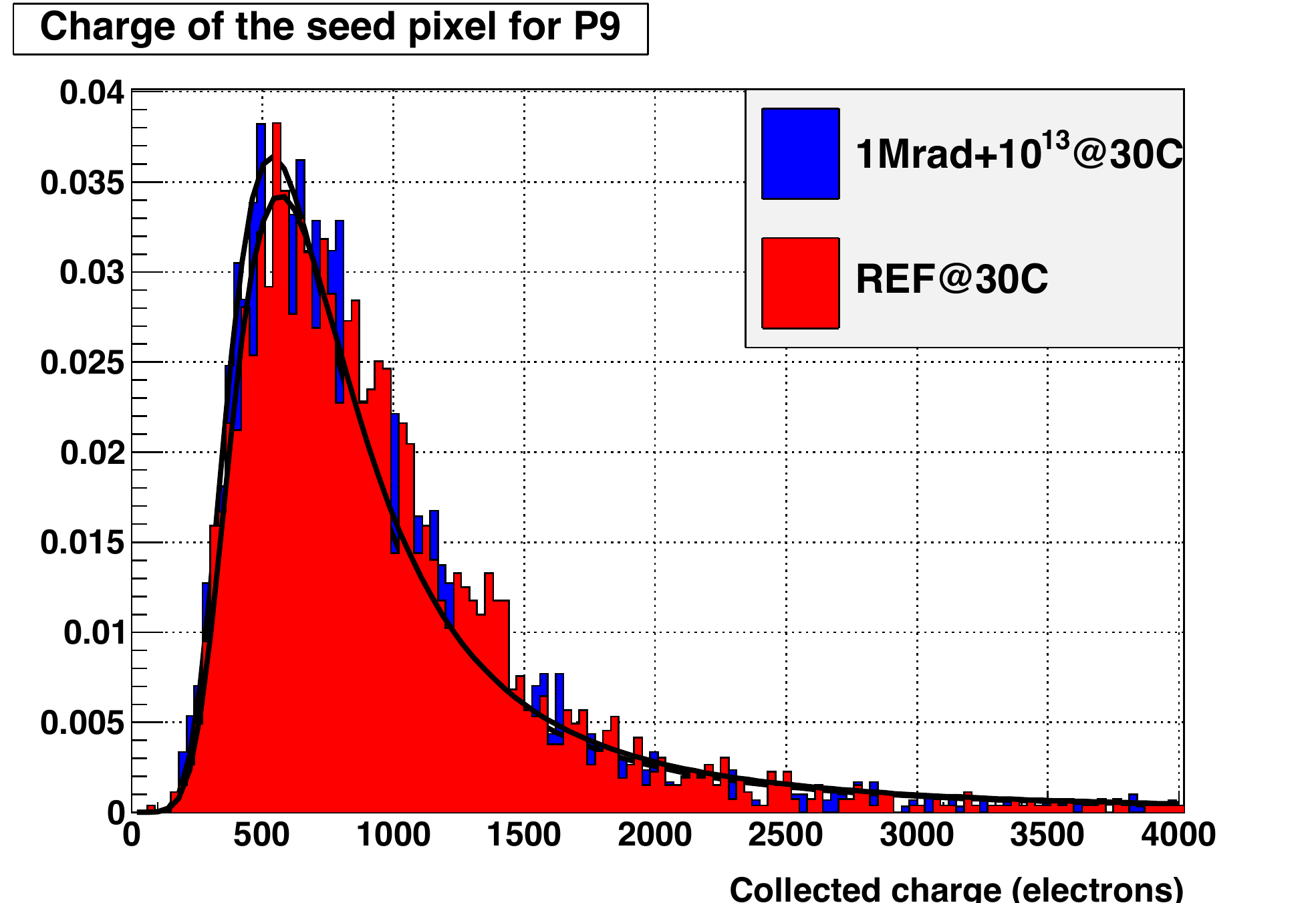}
\includegraphics[width=0.8\linewidth]{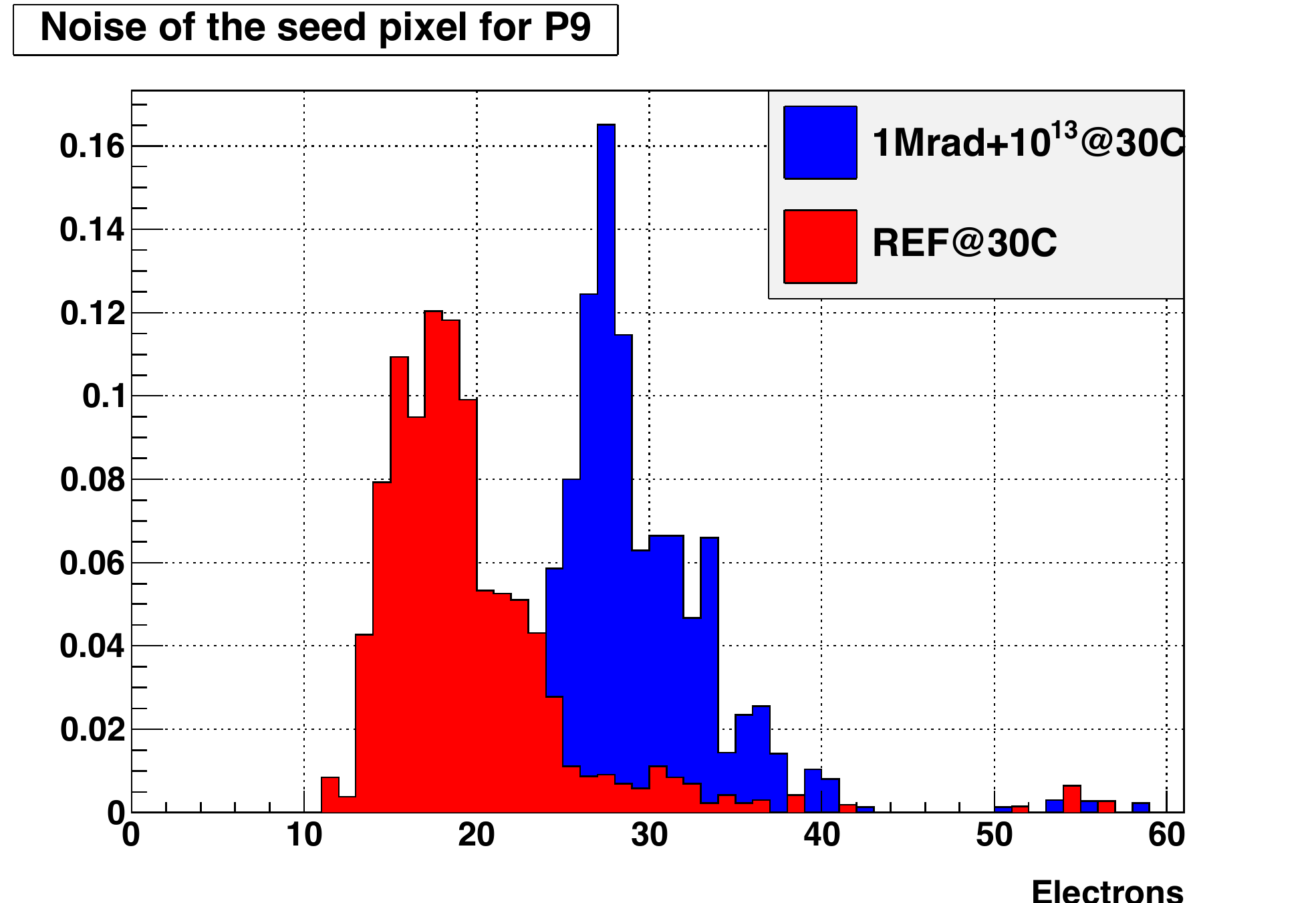}
\caption{MIMOSA32 deep pwell pixel beam test results: charge 
collected by the seed pixel of the reconstructed clusters (top)
and pixel noise (bottom) at T$_c$ = 30$^{\circ}$C.
The results are shown before (red) and after (blue) irradiating 
the sensor with 1~Mrad and 1~$\times$~10$^{13}$n$_{eq}$/cm$^2$.}
\label{fig:mimosa32-beam-P9-charge-noise}
\end{figure}

As an example, the hits reconstructed in the square pixels exhibited 
a typical cluster charge of $\sim$1100-1200~electrons, 
essentially concentrated in 2 to 4 pixels, about 40-50~\% 
of the charge being collected by the seed pixel. 
Figure~\ref{fig:mimosa32-beam-P9-charge-noise} displays 
the distribution of the charge collected by the cluster 
seed pixel in the deep pwell case. The measurements 
were performed at T$_c$ = 30$^{\circ}$C  
before and after irradiation. The radiation gives no observable effect
on the charge collection efficiency, but increases the noise by more than 50\%.   

The noise values observed reproduce those measured in the 
laboratory ($\lesssim$ 20~e$^-$ENC before irradiation and 
$\lesssim$ 30~e$^-$ENC after irradiation, at 
T$_c$ = 30$^{\circ}$C). The SNR varies accordingly 
from 30-35 (MPV) before irradiation, depending on the pixel 
variant, to 20-25 for the aforementioned combined radiation 
load. The detection efficiency, which is about 100\% before 
irradiation, remains nearly unchanged after irradiation (e.g. 
the deep pwell pixels exhibits still a detection efficiency 
of 99.87~$\pm$~0.07\%(stat)). 


Furthermore, an important aspect of the implementation will be the layout of the modules for the inner and outer 
pixel layers. First proposals for the data and power connection schemes have been worked out and are 
under investigation. Special emphasis is laid on the design and interconnection of a low mass bus cable
 which will provide power and data connections.

Progress has also been made in the development of hybrid pixel assemblies and strip detectors. 
Specifically, a production of thin hybrid assemblies with a total thickness of 150~$\mu$m has been 
demonstrated using dummy components. Furthermore, silicon strip sensors for the ALICE ITS upgrade 
have been designed and first prototypes have been produced, which will be used for module assembly 
and TAB-bonding tests.

The main objectives in the coming months are the completion of the radiation hardness studies 
of the TowerJazz structures and performance tests of pixel sensor prototypes, both in laboratory 
and test beam. Furthermore, the studies on the module layout and the data and power transmission
 scheme will be continued. 

\section{Mechanical layout and Cooling}

The design of the mechanical support structure and the services are driven by external constraints 
like required detector layout (layer grouping in three barrels, see Tab.~\ref{tab:itsNew}) and rapid 
accessibility for the installation. At the same time, inner constrains like sensor size (and choice),
 power dissipation and therefore cooling choice, and the total material budget have to be considered. 
A summary is given in Tab.~\ref{tab:fullITSRequirements}.

\begin{table}[h!]
\footnotesize
	\caption{Mechanical and Cooling requirements for the ITS upgrade}
	\centering
\begin{tabular}{|l|l|l|}
\hline
{Parameters} & { Inner Barrel}  & { Outer Barrel}  \\
\hline
\hline
{ Beampipe outer radius (mm)}	  &{ 20} &{ -} \\
{ Beampipe wall thickness (mm)}	  &{ 0.8} &{ -} \\
{ Detector Technology}		  &{ Pixel} &{ Pixel$-$Strip} \\						
{ Number layers}		&{ 3} &{ 4} \\
{ Mean radial positions (mm)}	&{ 22, 28, 36} &{ 200, 220, 410, 430} \\
{ Stave length in z (mm)}	&{ 270, 270, 270} &{ 843, 843, 1475, 1475} \\
{ Power consumption (W/cm$^2$)}	&{ 0.3~$\div$~0.5} &{ $\leq$ 0.5~mW/strip} \\
{ Material per layer ($\%$ of X$_0$)}&{ $\approx$~0.3} &{$\leq$ 1.0} \\
{ Working temperature ($^{\circ}$C)}	&{ $\approx$~30} &{ $\approx$~30}\\
\hline
\end{tabular}
	\label{tab:fullITSRequirements}
\end{table}

\begin{figure}[h!t]
    \centering
    \includegraphics[width=0.98\linewidth]{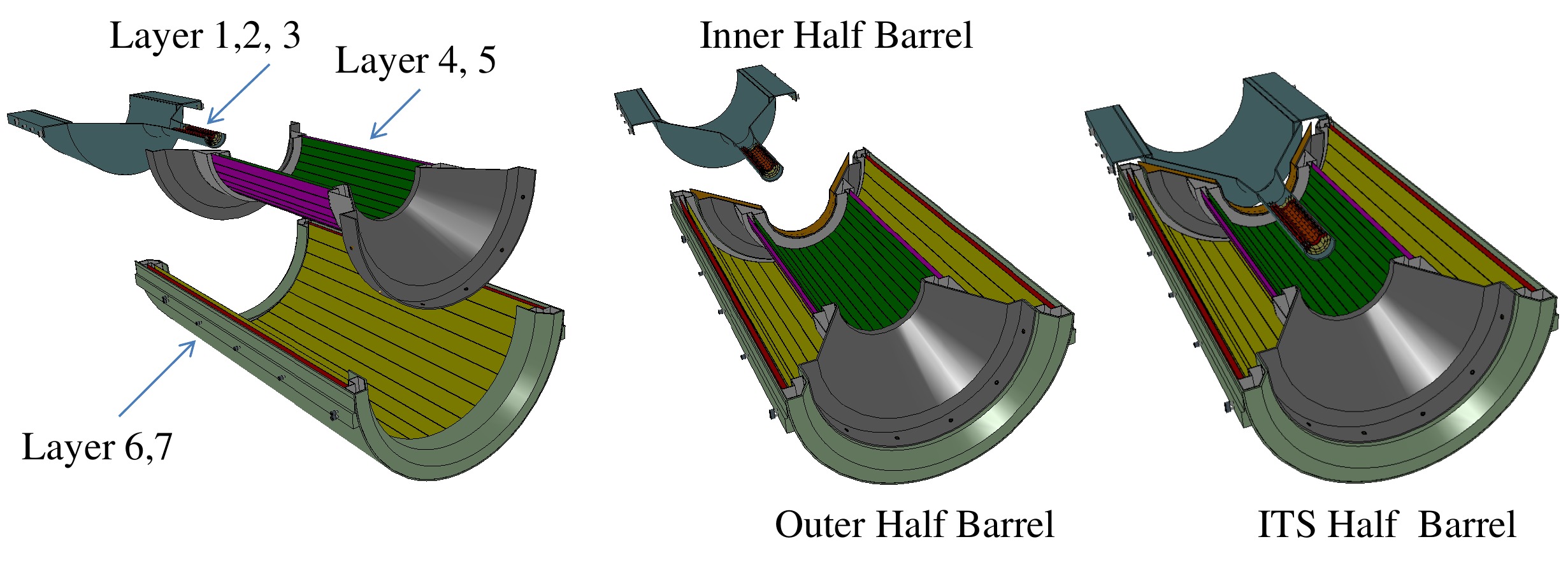}
    \caption{ITS modular mechanical structure.}
    \label{fig:integration2}
    \centering
    \includegraphics[width=0.98\linewidth]{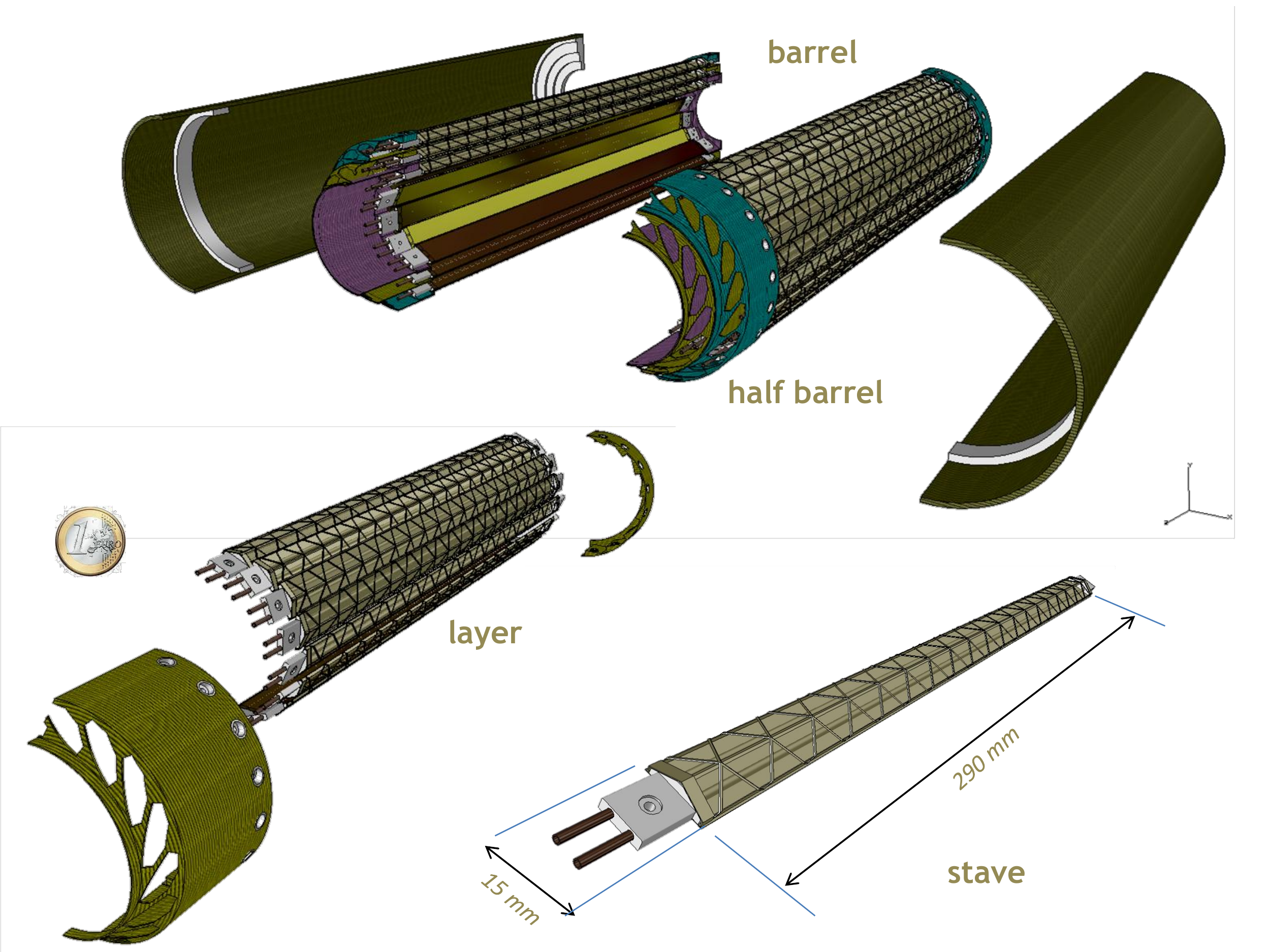}
    \caption{Drawings of the three innermost layers of the new ITS:
    from single stave to barrel.}
    \label{fig:sectorDraw}
\end{figure}

The ITS mechanical design is conceived as a two-barrel structure: Inner Barrel and Outer Barrel (see Fig.~\ref{fig:integration2}).
The design effort has been concentrated so far on the Inner Barrel due to the more stringent
requirements in terms of material budget.
The Inner Barrel has a modular structure consisting of 3 concentric layers, each segmented in the azimuthal
direction in modules referred to as "staves" (see Fig.~\ref{fig:sectorDraw}).

The stave is the smallest operable part of the detector. It is formed by 9 silicon pixel chips
(15\,mm~$\times$~30\,mm~$\times$~50\,$\mu$m thick) aligned in z to form a detection
area of 15\,mm~$\times$~270\,mm, which are flip-chip bump bonded on a flex-cable bus.
The stave includes a carbon fiber support structure, which provides a light and stiff frame for the 
sensors and  the electrical bus. Additionally this support integrates the cooling interface to remove the dissipated heat.\\

The different stave layouts which have been investigated can be divided in two
groups that refer to two different cooling concepts under study: {\itshape cold plates} and {\itshape cooling pipes}.

In the {\itshape cold plates} concept the cooling is provided by a  refrigerant fluid
(monophase or biphase) flowing in a microchannels structure, which is either in a
silicon (microchannel section area 0.04\,mm$^2$) or in a polyimide (microchannel section area 0.16\,mm$^2$)
substrate.The plate is in thermal contact with the front-end electronics while
the mechanical structure has to support the sensors with the bus and the {\itshape cold plates}.

In the {\itshape cooling pipes} option the cooling is provided by a refrigerant fluid (monophase or biphase)
flowing in 1.5\,mm outer diameter tube (0.035\,mm wall thickness). The tube is embedded in the mechanical structure
to which the sensors are glued.  Heat is transferred from the sensors to the pipes by the highly thermally
conductive structure.

Different prototypes of the mechanical structures for the two stave concepts have been produced, as for
example {\itshape Open Shell structures} as well as {\itshape Wound Truss Structure} (see~\cite[chap.\,5]{itsCDR} for details).
The prototypes were tested regarding their stiffness and weight, ease of assembly and handling.
The weight varies from 0.6\,g to 1.4\,g, for the structures produced from materials like carbon fiber T300 prepreg~\cite{toroycfa}
and M60J 3k (3000 filaments)~\cite{toroycfa}. Structural simulations and deformation tests resulted in a reasonable sag of
below 6~$\mu$m over a total length of 127~mm at twice the expected load factor. The first natural frequency was found at 596~Hz
which proves the stiffness of the structure.\\

Several cooling schemes fully integrated in the support structure and based on different refrigerant media (air, liquid and two-phase) have been studied. 

A solution using microchannels etched on a silicon substrate~\cite{mapelli} has been prototyped on a 4\,inch silicon wafer with
a thickness of 380\,$\mu$m and 0.1 - 0.5~$\Omega \cdot$cm p-type.
The microchannels are covered with a silicon layer 50\,$\mu$m thick.
The preliminary criteria adopted to choose the size of the silicon microchannels
(200\,$\mu$m\,$\times$\,200\,$\mu$m) are based on the studies published in~\cite{thome}.
The test performed so far, focused on thermal characterization of the single frame,
confirmed the high potentiality of this cooling concept when applied to the
ALICE requirements.

Another option is a Polyimide based Micro-Channels Heat Sink (MCHS)~\cite{MCHS}.
An analytical thermo-fluid dynamic study has been carried out to find the best microchannel section
(800\,$\mu$m\,$\times$\,200\,$\mu$m) fulfilling the operational requirements.
First prototypes were tested regarding their thermal behavior using PT100 sensors as well as thermographic imaging
and a kapton heater.
Tests revealed a uniformly increasing temperature along the flow direction.
In particular at a water flow rate of 2\,l/h and with a heat flux of 0.3\,W/cm$^2$,
the average temperature is less than 23$^\circ$C with a maximum value of 25.4$^\circ$C.

\begin{figure}[h]
    \centering
    \includegraphics[width=0.98\linewidth]{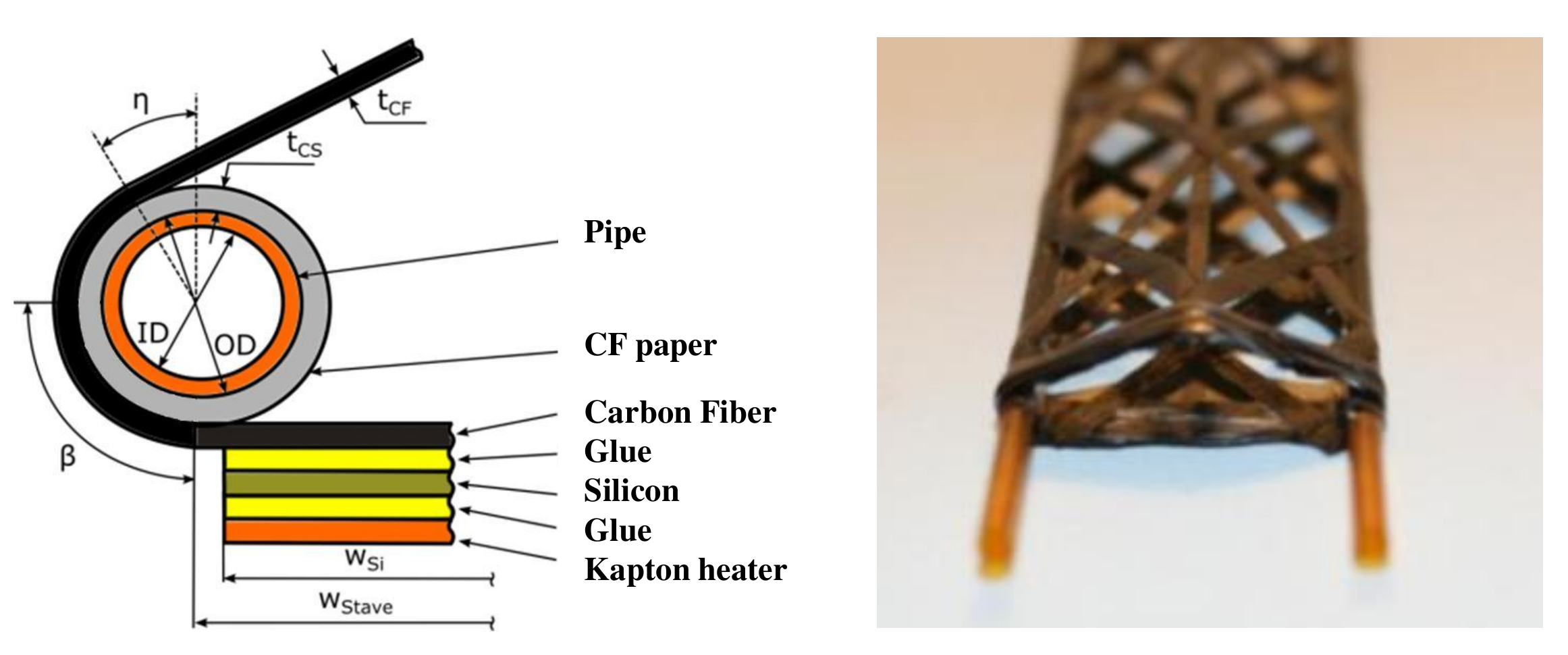}
    \caption{Picture of the wound truss structure with pipes
    (right) and schema of the section view (left).}
    \label{fig:D08prototype}
    \centering
    \includegraphics[width=0.98\linewidth]{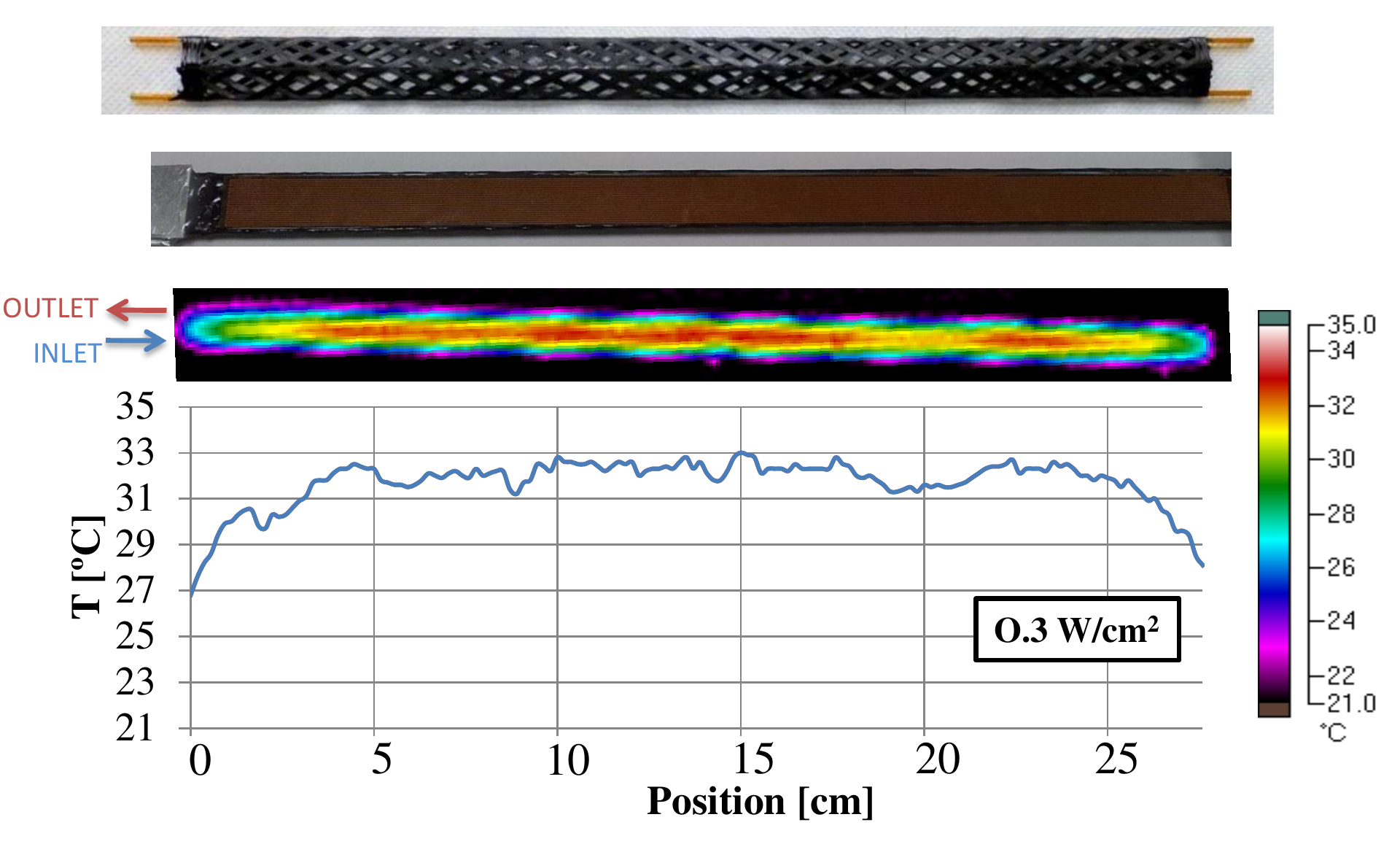}
    \caption{Thermographic images of the heater surface with water flow rate of 8\,l/h
    and power consumption of 0.3\,W/cm$^2$. The temperature profile is taken along the center the structure.}
    \label{fig:D08ThermalTest}
\end{figure}

A third option is the usage of cooling pipes suitable for either single phase or evaporative cooling.
Two different prototypes have been tested: 
a) Wound truss structure with pipes and uniform carbon plate b) Wound truss structure with only pipes.\\
As an example, option b) is shown in Fig~\ref{fig:D08ThermalTest} where the thermal contact is enhanced
by a graphite foil wrapped all around the pipes.
The thermal results obtained with option b) are shown in Figure~\ref{fig:D08ThermalTest}.
The test has been carried out using the flow rate of 8\,l/h with the water temperature at the inlet
of 14.8$^\circ$C. The power density applied is 0.3\,W/cm$^2$. The average temperature is less than 
31$^\circ$C with a maximum value of 33$^\circ$C.

The material budget estimated for the stave prototypes as mentioned above fulfill the mechanical 
and thermal requirements of the new ITS, although further studies and optimization are still needed.
All prototypes results in a mean value of the material budget below 0.31\,X/X$_0$.
 
As an example, the material budget of the prototype using carbon fiber wound structure with embedded polyimide tubes
are reported in Tab.~\ref{tab:materialBudget1}. It accounts for the percentage of surface covered by both top and bottom carbon fibers.
The wall of the polyimide tube is 35\,$\mu$m thick and the water has been considered as coolant.
In addition we accounted for 2 layers of glue 100\,$\mu$m thick assuming a coverage of 50$\%$
for the layer between the structure and the silicon.

The mean value for this prototype is approximately 0.26\,X/X$_0$.
It is important to point out that this structure is slightly lighter than the other options mentioned above,
but it has reduced thermal performance.
The R$\&$D activity on this structure is important because it will allow us to further 
reduce the material budget if a low power density ($\le$0.3\,W/cm$^2$) chip is chosen for the final design.

\begin{table}[h!]
\footnotesize
\caption{Wound truss structure with pipe prototype: list of components and their
	contribution to the mean material budget estimate.}
\begin{tabular}{|l|c|c|c|c|c|}
\hline
\multirow{2}{*}{Material} & {Surface} & {Thickness}  & {X$_0$ } & {X/X$_0$ } & \multirow{2}{*}{x ($\%$) $^{*)}$}\\
                 & {($\%$)} & {($\mu$m)}		  & {(cm)} & {($\%$)} & {}\\
\hline
\hline
{ CFRP filament} & { 100} &{ 70}   &{ 25} &{ 0.035}&{ 12.6}\\
{ Polyimide Tubes} & { 19} &{ 70}  &{ 28.6} &{ 0.005}&{ 1.8}\\
{ Water} & { 19} &{ 1450}	  &{ 36.1} &{ 0.06}&{ 22.8}\\
{ Glue (CFRP - silicon)} & { 50} &{ 100}  &{ 44.4} &{ 0.01}&{ 4.4}\\
{ Silicon} & { 100} &{ 50}	  &{ 9.36} &{ 0.054}&{ 20.7}\\
{ Glue (silicon - bus)} & { 100} &{ 100} &{ 44.4} &{ 0.022}&{ 8.7}\\
{ Electrical bus} & { 100} &{ -}  &{ -} &{ 0.075}&{ 29.0}\\
\hline
\hline
\multicolumn{1}{|l}{ Total} & \multicolumn{1}{c}{ } &\multicolumn{1}{c}{ }
						  &\multicolumn{1}{c}{ } &\multicolumn{1}{c}{ $\approx$ 0.26}&{ }\\				
\hline
\end{tabular}
$^{*)}$ Contribution to the total X/X$_0$ ($\%$)
	\label{tab:materialBudget1}
\end{table}

More details regarding the planned mechanical layout, the material budget contributors, the services, the integration procedure
 and the different cooling options can be found in ~\cite[chap.\,5]{itsCDR}.

\section{Time Schedule}

Although the ALICE ITS upgrade schedule is not linked directly to the LHC upgrade and its schedule, the installation of a new ITS will require a long shutdown (LS) and, therefore, will naturally have to be in phase with the installation of upgrades for the other LHC experiments, planned as of today for the 2013/14 and 2017/2018 shutdowns.

In this respect, the ALICE ITS upgrade will use the time between 2012 and 2014 for the R\&D phase, for prototyping and for testing.
 The production and pre-commissioning phase will start in 2015. The commissioning and the installation is planned for the long LHC shutdown period in 2017/2018.
Details are given in Tab.~\ref{tab:timeITS}.

\begin{table}[htp]
  \small
  \centering
  \caption{ITS upgrade timeline ~\cite{ALICE_LOI}}
   \begin{tabular}{p{1.6cm}p{6.4cm}}
    \hline\hline
      \textbf{Year} & \textbf{Activity} \\
    \hline\hline
     {2012 -- 2014} &  \textbf{R\&D} \\
     \hline
     2012 & finalization of detector specifications; evaluation of detector technologies (radiation and beam tests);
           first prototypes of sensors, ASICS, and ladders (demonstrators);  \\
    2013 & selection of technologies and full validation; engineering design for sensors, ASICs, ladders, data links; 
            engineering design for support mechanics and services; Technical Design Report; \\
    2014 & final design and validation \\
    \hline
    {2015 -- 2018} &  \textbf{Construction and Installation} \\
    \hline
     {2015 -- 2016} & {production, construction and test of detector modules} \\
    {2017}    & {assembly and pre-commissioning in clean room} \\
    {2018} & {installation in the cavern} \\
   \hline\hline
    \end{tabular}%
  \label{tab:timeITS}%
\end{table}%

\section*{Acknowledgment}
The content of this contribution represent only a superficial overview on the studies unveiled in the Conceptual Design Report~\cite{itsCDR}. Therefore, I would like to express my gratitude to the various contributors of the CDR.
\vspace*{1cm}




\bibliographystyle{elsarticle-num}
\bibliography{refs}


\end{document}